\documentclass[12pt]{article}
\usepackage{epsfig}
\newcommand{\bea}{\begin{eqnarray}}
\newcommand{\eea}{\end{eqnarray}}
\newcommand{\be}{\begin{equation}}
\newcommand{\ee}{\end{equation}}

\begin{document}

\setcounter{page}{0} \thispagestyle{empty}

\vskip 2cm

\begin{center}
{\Large \textbf{Pomeron in the ${\mathcal N}=4$ supersymmetric gauge
model at strong
couplings}} \\[5mm]
\vspace*{10mm} A.~V.~Kotikov$^{a}$, L.~N.~Lipatov$^{b,c}$,
\\[10mm]

${}^{a}$ Bogoliubov Laboratory of Theoretical Physics \\[0pt]
Joint Institute for Nuclear Research\\[0pt]
141980 Dubna, Russia \\[0pt]
\vspace*{0.5cm} ${}^{b}$ II. Institut f\" ur Theoretische Physik, \\[0pt]
Universit\" at Hamburg\\[0pt]
Luruper Chaussee 149, \\[0pt]
22761 Hamburg, Germany \\[0pt]
\vspace*{0.5cm} ${}^{c}$ Theoretical Physics Department, \\[0pt]
Petersburg Nuclear Physics Institute \\[0pt]
Orlova Rosha, Gatchina, \\[0pt]
188300, St. Petersburg, Russia\\[0pt]
\end{center}

\vspace{1cm}

\begin{center}
\textbf{Abstract}
\end{center}

We find the BFKL Pomeron intercept at ${\mathcal N}=4$ super-symmetric gauge theory in the form of the 
inverse coupling expansion 
$j_0=2-2\lambda^{-1/2}-\lambda^{-1}
+ 1/4 \, \lambda^{-3/2} + 2(1+3\zeta_3)\lambda^{-2}
+ O(\lambda^{-5/2})$ 
with the use of the AdS/CFT correspondence in terms of string energies calculated recently.
The corresponding slope $\gamma'(2)$ of the anomalous dimension
calculated directly up to the fifth order of perturbation theory turns out to be in
an agreement with the closed expression
obtained from the recent Basso results.
\\


\newpage
\vskip 2cm

\section{Introduction}
\indent

Pomeron is the Regge singularity of the $t$-channel partial wave
%initially proposed by Chew, Frautschi and Gribov
\cite{Chew:1961ev}
%even before Quantum Chromodynamics (QCD) was introduced. It is
responsible for the approximate
%possible
equality of total cross-sections
for high energy particle-particle and particle-antiparticle interactions 
valid in an accordance with the Pomeranchuck theorem\cite{Pomeranchuk}.
In QCD
the Pomeron
is a
%coherent color-singlet
colorless object, constructed
%which is built
from reggeized gluons \cite{BFKL}.

The investigation of the high energy behavior of scattering amplitudes in the
${\mathcal N}=4$ Supersymmetric
Yang-Mills (SYM) model \cite{KL00,KL,Fadin:2007xy} is important for
our
understanding of the Regge processes in QCD. Indeed, this conformal model
can be considered as a simplified version of QCD, in which
the next-to-leading order (NLO) corrections \cite{next} to the
Balitsky-Fadin-Kuraev-Lipatov (BFKL) equation \cite{BFKL}
are comparatively simple and numerically
small. In the ${\mathcal N}=4$ SYM the equations for composite states of several
reggeized gluons and for anomalous dimensions of quasi-partonic
operators turn out to be integrable at the leading logarithmic approximation \cite{L93, L97}.
Further, the eigenvalue of the BFKL kernel for this model 
%here
has the remarkable 
%last fact is related to the a
property of
the maximal transcendentality
%valid in the perurbation theory for the ${\mathcal N}=4$ SYM model
\cite{KL}. This property
%of ${\mathcal N}=4$ SYM
gave a possibility to calculate the
anomalous dimensions (AD) $\gamma$ of the twist-2 Wilson operators in one
\cite{Lipatov00},
two \cite{KL,Kotikov:2003fb}, three \cite{Kotikov:2004er}, four
\cite{Kotikov:2007cy,Janik} and five \cite{Lukowski:2009ce} loops using
the QCD results \cite{Moch:2004pa}
and the asymptotic  Bethe ansatz \cite{Beisert:2005fw} improved with
wrapping corrections \cite{Janik}
\footnote{The
%exact
anomalous dimensions up to four loops were calculated also with the use of
the Baxter equation \cite{Kotikov:2008pv}.}
in an agreement with the BFKL predictions  \cite{KL00,KL}.

On the other hand, due to the AdS/CFT-correspondence
\cite{AdS-CFT,AdS-CFT1,Witten}, in ${\mathcal N}=4$ SYM
some physical quantities can be also computed
at large couplings.
In particular, for AD of the large spin operators 
Beisert, Eden and Staudacher constructed
%results are even more complete:
the integral equation \cite{Beisert:2006ez} with the use
the asymptotic
Bethe-ansatz. This equation reproduced the known results
%large spin case
at small coupling constants
%being also
and
%it
is in a full agreement (see \cite{Benna:2006nd, Basso:2007wd}) with large
coupling predictions
\cite{Gubser:2002tv, Frolov:2002av}.

With the use of the BFKL equation in a diffusion approximation
%\cite{KL}-\cite{BFKL},
\cite{BFKL,KL00,Fadin:2007xy},
strong coupling results
%expressions
for AD
\cite{Gubser:2002tv} and the pomeron-graviton duality \cite{Polchinski:2002jw}
%several years ago
the Pomeron intercept was calculated
at the leading order in the inverse
%large
coupling constant (see the Erratum\cite{Kotikov:2006}
to the paper \cite{Kotikov:2004er}).
\footnote{The value of this intercept was estimated earlier in Ref.\cite{Kotikov:2003fb}.}
Similar results in the   ${\mathcal N}=4$ SYM and QCD were obtained in
Refs. \cite{Brower:2006ea} and \cite{Stasto:2007uv}.
The Pomeron-graviton duality in the   ${\mathcal N}=4$ SYM
gives a possibility to construct the Pomeron interaction model
as a generally covariant effective theory for the reggeized gravitons
\cite{Lipatov:2011ab}.

Below we use  recent
%string
calculations \cite{Gromov:2011de,Basso:2011rs,Gromov:2011bz,Roiban:2011fe}
of string energies to
%, we and
find
%in the strong coupling expansion the second and two
the
%next order
strong coupling
corrections
to the Pomeron intercept $j_0=2-\Delta$ in next orders.
We discuss also the relation between the Pomeron intercept and the slope
%, perturbative expansion
of the anomalous dimension at $j=2$.

\section{BFKL  equation at small coupling constant}
\indent

The eigenvalue of the BFKL equation in  ${\mathcal N}=4$ SYM model has
the following perturbative expansion
%in the perturbation theory the following form
\cite{KL00,KL}
(see also Ref.
\cite{Fadin:2007xy})
\be
j-1 = \omega = \frac{\lambda}{4\pi^2} \biggl[\chi(\gamma_{BFKL}) +
\delta(\gamma_{BFKL})
\frac{\lambda}{16\pi^2}\biggr], ~~~ \lambda=g^2N_c,
\label{1i}
\ee
where $\lambda$ is the t'Hooft coupling constant. 
%Generally the coefficients 
The quantities $\chi$ and $\delta$
are functions of
the conformal weights
$m$ and $\widetilde{m}$ of the principal series of unitary M\"{o}bius group representations, but for the
conformal spin $n=m-\widetilde{m}=0$ they depend only on the BFKL anomalous dimension
\be
\gamma_{BFKL}=\frac{m+\widetilde{m}}{2}=\frac{1}{2}+i\nu \,
\label{2i}
\ee
and are presented below \cite{KL00,KL}
%$\lambda=g^2 N$
\bea
\chi(\gamma) &=& 2\Psi(1)-\Psi(\gamma)-\Psi(1-\gamma), \label{2i} \\
\delta(\gamma) &=& \Psi^{''}(\gamma)+\Psi^{''}(1-\gamma) + 6\zeta_3
-2\zeta_2 \chi(\gamma) -2\Phi(\gamma)-2\Phi(1-\gamma) \, .
\label{3i}
\eea

Here $\Psi (z)$ and $\Psi ^{\prime }(z)$, $\Psi ^{\prime \prime }(z)$ are
the Euler $\Psi $ -function and its derivatives.
%and $a=g^{2}N_{c}/(16\pi ^{2})$ is coupling constant in DREG scheme.
The function $\Phi(\gamma)$ is defined as follows
%given
\be
\Phi(\gamma) = 2 \sum_{k=0}^{\infty }
\frac{1}{k+\gamma} \, \beta ^{\prime }(k+1)\,,
\label{4i}
\ee
where
\be
\beta ^{\prime }(z)=\frac{1}{4}\Biggl[
\Psi ^{\prime }\Bigl(\frac{z+1}{2}\Bigr)-
\Psi ^{\prime }\Bigl(\frac{z}{2}\Bigr)\Biggr]\,.
\label{5i}
\ee

Due to the symmetry of $\omega$ to the substitution
$\gamma _{BFKL}\rightarrow 1-\gamma _{BFKL}$ expression (\ref{1i}) is an
%the
even function of $\nu$
\be
\omega = \omega_0 + \sum_{m=1}^{\infty} (-1)^m \, D_m \, \nu^{2m} \, ,
\label{6i}
\ee
where
\bea
 \omega_0 &=& 4\ln 2 \, \frac{\lambda}{4\pi^2} \left[ 1- \overline{c}_1
\frac{\lambda}{16\pi^2}  \right] +
O(\lambda^3) \,, \label{7i} \\
D_m &=&
%(-1)^m \,
2\left(2^{2m+1}-1\right)\zeta_{2m+1} \frac{\lambda}{4\pi^2} +\frac{\delta ^{(2m)}(1/2)}{(2m)!}\,
\frac{\lambda ^2}{64 \pi ^4}+
O(\lambda^3) \,.
\label{8i}
\eea
According to Ref. \cite{KL} we have
\be
%\hat{\delta}_1
 \overline{c}_1 ~=~ 2 \zeta_2 +
\frac{1}{2\ln 2} \left(11\zeta_3
- 32{\rm Ls}_{3}\Bigl(\frac{\pi }{2}\Bigl) -14 \pi \zeta_2
\right) \approx  7.5812 \,,
\label{8.1i}
\ee
where (see \cite{Lewin})
%\cite{Lewin,Devoto})
\be
{\rm Ls}_{3}(x)=-\int_{0}^{x}\ln ^{2}\left| 2\sin \Bigl(\frac{y}{2}%
\Bigr)\right| dy \,. \label{8i}
\ee

Thus, the rightmost
%leading
Pomeron singularity of the partial wave $f_j(t)$
%$j$
in the perturbation theory
is situated at
%with
\be
j_0 = 1+\omega_0 = 1+ 4\ln 2 \, \frac{\lambda}{4\pi^2} \left[ 1- \overline{c}_1
\frac{\lambda}{16\pi^2}  \right] +
O(\lambda^3) \label{10i}
\ee
for small values of coupling $\lambda$.
%Note
In turn, the anomalous dimension $\gamma$ also has the square root singularity in this point,
which means, that the convergency radius
of the  perturbation series in $\lambda$
for the anomalous dimension
$\gamma=\gamma(\omega,\lambda)$ at small $\omega$ is given by the expression
\be
\lambda_{cr}=
\frac{\pi^2\omega}{\ln 2} \left( 1 + \overline{c}_1 \frac{\omega}{16\ln 2}\right)
+ O\Bigl(\omega^3\Bigr) \,.
\ee

%{\bf BFKL intercept and the convergency of PT series for $\gamma$}

Let us write the expression (\ref{6i}) for $\omega$ in the diffusion approximation for arbitrary $\lambda$
\be
\omega =\omega_0(\lambda )-D_1(\lambda )\,\nu ^2+O(\nu ^4)\,.
\ee
From this result we obtain, that the anomalous dimension has the square-root singularity
\be
\lim _{\lambda \rightarrow \lambda _{cr}}\gamma =
\sqrt {\frac{\omega'_0(\lambda_{cr} )\,(\lambda_{cr}-\lambda )}{D_1(\lambda_{cr} )}} \,,
\ee
where $\lambda_{cr}$ is a function of $\omega$ satisfying the equation
\be
\omega =\omega_0(\lambda _{cr})\,.
\ee
Therefore the perturbative series for the anomalous dimension $\gamma$ 
\be
\gamma =\sum _{k=1}^\infty \lambda ^k\,c_k(\omega)\,,
\ee
has the finite radius divergency $\lambda =\lambda_{cr}$ and its coefficients $c_k$ behave
 at large $k$ as follows
\be
\lim _{k\rightarrow \infty }c_k= \lambda_{cr}^{-k-\frac{3}{2}}\,\frac{1}{2\sqrt{\pi}}\,
\sqrt {\frac{\lambda _{cr}\omega'_0(\lambda_{cr} )}{D_1(\lambda_{cr} )}}\,.
\ee

It will be interesting to find higher order corrections to the BFKL intercept $\omega _0 (\lambda)$ 
and the diffusion coefficient $D_1(\lambda)$ by comparing the above asymptotic expression for $c_k$ with the analytic
results at $k=1-5$ obtained recently
%It will %would
%be interesting %important
%to find higher order corrections to the BFKL intercept $j_0$ %this formula
%from the investigation of convergency of the perturbation theory %in $\lambda$ 
%using the analytic results for the anomalous dimensions obtained recently
\cite{KL,Kotikov:2004er,Kotikov:2007cy,Janik,Lukowski:2009ce}.
Note, that the BFKL singularity for positive $\omega$ is situated
at positive $\lambda = \lambda_{cr}$. But it is expected, that
with growing $\omega$ the nearest singularity, responsible for
the perturbation theory divergency will be at negative
$\lambda$. Positions of both singularities can be
found from the perturbative expansion of $\gamma$ with the
possible use
of appropriate resummation methods (cf. \cite{Kotikov:2004er}).

Due to the M\"{o}bius invariance and hermicity of the BFKL hamiltonian in ${\mathcal N}=4$ SUSY
%the the 
expansion (\ref{6i})
%representation (\ref{9i})
is valid also at large coupling constants.
%applicable.
In the framework of the AdS/CFT correspondence the BFKL Pomeron is equivalent to
%coincides with
the reggeized graviton
~\cite{Polchinski:2002jw}.
%which means, that
In particular,
in the strong coupling regime $\lambda \rightarrow \infty$
\begin{equation}
j_0~=~ 2-\Delta \,,
\label{11i}
\end{equation}
where the leading contribution
%term 
$\Delta =
%\sim
2/\sqrt{\lambda}$ was
%a small number
calculated in Refs.~\cite{Kotikov:2006, Brower:2006ea,
Stasto:2007uv}. Below we find next-to-leading terms in
%corrections to 
the strong coupling expansion
of the Pomeron intercept. 
%But 
In the next section the simple approach to the intercept estimates discussed shortly in 
Ref.~\cite{Kotikov:2006}
will be reviewed.

\section{ AdS/CFT correspondence}
%\section{Energy-momentum conservation}

\indent

Due to the energy-momentum conservation, the universal anomalous dimension of the
stress tensor $T_{\mu \nu}$ should be zero, i.e.,
\be
\gamma(j=2)=0.
\label{1e}
\ee

It is important, that the anomalous dimension $\gamma$
contributing to the DGLAP
equation \cite{DGLAP}
does not coincide with $\gamma_{BFKL}$ appearing in the BFKL equation.
They are related as follows \cite{next} (see also
%nice review in
\cite{Salam:1998tj})
\be
\gamma ~=~ \gamma_{BFKL} + \frac{\omega}{2} ~=~ \frac{j}{2}+i\nu \,,
%\gamma_{BFKL} + \frac{\omega}{2} ~=~ \gamma_{BFKL} + \frac{j-1}{2},
\label{2e}
\ee
where the additional contribution $\omega /2$ is responsible in particular for the cancelation
of the singular terms
$\sim 1/\gamma ^3$  obtained from the NLO
corrections (\ref{1i}) to the eigenvalue of the BFKL kernel
%equation
\cite{next}.

Using  above relations
%sum  rule
one obtains
\be
\nu(j=2)=i\,.
\label{4e}
\ee
As a result, from eq. (\ref{6i}) for the Pomeron intercept
%and (\ref{11i})
we derive the following
representation for the correction
%parameter
$\Delta$  (\ref{11i})
%in eq. (\ref{11i})
to the graviton spin $2$
\be
\Delta ~=~  \sum_{m=1}^{\infty} D_m .
\label{5e}
\ee

In the diffusion approximation,
where $D_m=0$ for $m\geq 2$,
one obtains from (\ref{5e}) the relation between  the diffusion coefficient
$D_1$ and $\Delta$ (see ~\cite{Kotikov:2006})
%Erratum to the paper ~\cite{Kotikov:2004er})
\be
D_1 ~\approx ~ \Delta \, .
\label{6e}
\ee
This relation was also obtained in Ref. \cite{Brower:2007xg}.
%they confirmed relation (\ref{6e}) in the diffusion approximation.\

According to
%(\ref{9i}),
(\ref{11i}) and (\ref{5e}), we have the following
small-$\nu$ expansion for the eigenvalue of the BFKL kernel
%position of the $j$-plane singularities
\be
j-2 = \sum_{m=1}^{\infty}  D_m \left({(-\nu^{2})}^m-1\right),
\label{7e}
\ee
where $\nu^2$ is related to $\gamma$ according to eq. (\ref{2e})
\be
\nu^2=
-{\left(\frac{j}{2}-\gamma\right)}^2.
\label{8e}
\ee

On the other hand, due to the
 ADS/CFT correspondence the string energies $E$ in dimensionless units are related to the anomalous dimensions
$\gamma$ of the twist-two
%local 
operators as follows
\cite{AdS-CFT,Gubser:2002tv}\footnote{Note 
that our definition (\ref{1a}) of the  string energy $E$ differs from the standart one, where
 $E$ is equal to the scaling dimension $\Delta_{sc}$. The eq. (\ref{1a}) can be represented as $E^2=\Delta_{sc}^2-J^2$,
where $J$ is twist. Our definition (\ref{1a}) is convenient partially for the large $\lambda$ expansion (\ref{5a})
which started with $J^2$ for  $\Delta_{sc}^2$ (see, for example, Ref. \cite{Gromov:2011bz} and discussion and references
therein).}

\begin{equation}
E^2=(j+\Gamma)^2-4,~~\Gamma=-2\gamma
\label{1a}
\end{equation}
and therefore we can obtain from (\ref{8e}) the relation between the parameter $\nu$ for
the principal series of unitary representations of the M\"{o}bius group and the string
energy $E$
\be
\nu^2=
-\left(\frac{E^2}{4}+1\right)\,.
\label{3a}
\ee
This expression for $\nu ^2$ can
%should
be inserted in the r.h.s. of Eq. (\ref{7e}) leading to the following expression for the Regge trajectory
of the graviton in the anti-de-Sitter space
\be
j-2 =  \sum_{m=1}^{\infty}
D_m \left[{\left(\frac{E^2}{4}+1\right)}^m-1\right].
\label{4a}
\ee

 Note \cite{Kotikov:2006}, that due to (\ref{3a})
%we obtain,
expression (\ref{6i}) for
the eigenvalue of the BFKL kernel in the diffusion approximation
%(\ref{6i}) with $D_m=0$ for $m\geq 2$, i.e.
(\ref{6e})
\be
j ~=~ j_0 - \Delta \nu^{2} ~=~ 2  - \Delta \Bigl(\nu^{2} +1 \Bigr) \, ,
\label{9ia}
\ee
is equivalent to the linear
%expression for the
graviton Regge trajectory
\begin{equation}
j=
%2 +\frac{\alpha'}{2}t\,,\,\,\alpha' t=\Delta \,E^2/2\,,
2 + \frac{\alpha'}{2}t\,,\,\,\alpha' t= \Delta \, \frac{E^2}{2}\,,
%\frac{\alpha'}{2}t=\Delta E^2,~~ \Delta t=E^2/R^2,\,
\label{2a}
\end{equation}
where its slope $\alpha'$ and the Mandelstam invariant $t$, defined in the $10$-dimensional space, equal
\be
\alpha' ~=~ \Delta \,\frac{R^2}{2},~~~ t ~=~ \frac{E^2}{R^2}
\label{2ab}
\ee
and $R$ is the radius of the anti-de-Sitter space.

%Considering together the equations (\ref{8e}) and (\ref{1a}),

Now we return to the eq.
%general expression
(\ref{4a}) in general case. We
%shall
assume below, that it is valid also
%in the region of
at large $j$ and large $\lambda$ in the region
\be
1\ll j \ll\sqrt{\lambda}\,,
\label{jll1}
\ee
where the strong coupling calculations of energies
%anomalous dimensions
were performed~\cite{Gromov:2011de,Roiban:2011fe}.
Comparing the l.h.s. and r.h.s. of (\ref{4a}) at large $j$ values
gives us the
%results for the
coefficients $D_m$ and $\Delta$ (see Appendix A).
\footnote{ {When this paper was almost prepared for publication, we found the article
\cite{Costa:2012cb} containing some of our results
(see discussions in  Appendix A).}}

\section{Graviton Regge trajectory and Pomeron intercept
}

The coefficients $D_1$ and $D_2$ at large $\lambda$ can be written as follows
%values
\footnote{ Here we consider only the calculation of the
$\lambda^{-1}$ correction to Pomeron intercept. More general results are
presented in Appendix A.}
\be
D_1 ~=~ \frac{2}{\sqrt{\lambda}} \left(1-
\frac{2a_{01}}{\sqrt{\lambda}}\right),~~
D_2 ~=~ -\frac{8 a_{10}}{\lambda^{3/2}}\,,
%8\ln 2 -%3 \frac{7}{2}
 \label{Li1}
\ee
%obtained
where $a_{01}$ and $a_{10}$ are calculated
in Appendix A
\be
a_{01}
= -\frac{1}{4},~~
 a_{10} =  \frac{3}{8}\,.
\ee
As a result,
we find eigenvalue (\ref{4a}) of the BFKL kernel at large $\lambda$
%. We obtain
in the form of the nonlinear Regge trajectory of the graviton in the anti-de-Sitter space
\be
j-2 ~=~ D_1 \, \frac{E^2}{4} + D_2 \, \left[ \left(\frac{E^2}{4}\right)^2
+ \frac{E^2}{2}\right], 
%~~~ E^2=2\alpha' t 
\, .
 \label{Li2}
\ee

Note, that the
%The
perturbation theory for the BFKL equation gives this trajectory at small
$\omega=j-1$ (see eq. (\ref{1i})) according to
%, where
%\[
%\gamma_{BFKL}=\frac{1}{2}+i\nu,~~ \nu^2 ~=~ -\left(\frac{E^2}{4} +1\right) \,,
%%\nonumber 
%%\label{Li3}
%\]
%%{\bf 
%as it was shown in 
eqs. (\ref{2e}) and (\ref{3a}).
%}
However the energy-momentum constraint (\ref{1e}), leading to $\omega =1$ at $E=0$,
is not fulfilled in the
perturbation theory, because at $\gamma \to 0$ the right-hand side of
(\ref{1i}) contains the pole singularities which should be cancelled after an
appropriate resummation of
%an appropriate way.
all orders.

Neglecting
%in (\ref{Li2})
the term $D_2 E^2/2 ~\sim~ E^2/\lambda^{3/2}$ at $\lambda \to \infty$
 in
comparison with a larger correction $a_{01} E^2/\lambda$, we
obtain the graviton trajectory (\ref{Li2}) in the form
\be
j-2 ~=~ \frac{2}{\sqrt{\lambda}} \left(1-
\frac{2a_{01}}{\sqrt{\lambda}}\right) \,
\frac{E^2}{4} - \frac{8a_{10}}{\lambda^{3/2}} \, \left(\frac{E^2}{4}\right)^2 \, .
 \label{Li4}
\ee

Solving this quadratic equation, one can derive with
%we obtain
the same accuracy (see \cite{Gromov:2011de,Roiban:2011fe})
\be
\frac{2}{\sqrt{\lambda}} \, \frac{E^2}{4} ~=~ \Bigl(j-2\Bigr) \, \left( 1+
 2\frac{a_{01}+a_{10}(j-2)}{\sqrt{\lambda}}\right) \,.
 \label{Li5}
\ee

On the other hand, due
%$E^2=(j-2\gamma)^2-4$ according
to (\ref{1a}) this relation can be written as follows
%Therefore
\be
\frac{1}{2\sqrt{\lambda}} \, \Bigr(j-2\gamma\Bigr)^2 ~=~
\frac{2}{\sqrt{\lambda}} + \Bigl(j-2\Bigr) \, \left( 1+
 2\frac{a_{01}+a_{10}(j-2)}{\sqrt{\lambda}}\right)
 \label{Li5}
\ee
and for $j-2 >> 1/\sqrt{\lambda}$ we have
%$j\neq 2$
\be
j-2\gamma ~=~ \sqrt{2(j-2)} \, \lambda^{1/4} \, \left[
1+\left(\frac{1}{j-2}+a_{01}+a_{10}(j-2)\right)
\, \frac{1}{\sqrt{\lambda}}\right]\,.
 \label{Li6}
\ee

In particular, for $j=4$ one obtains the anomalous dimension for
the Konishi operator $\gamma=\gamma_K$ \cite{Gromov:2011de}
(see also Appendix B)
\be
2-\gamma_K ~=~ \lambda^{1/4} \, \left[
1+\left(\frac{1}{2}+a_{01}+2a_{10} \right)
\, \frac{1}{\sqrt{\lambda}}\right] =
\lambda^{1/4} \, \left[
1+ \frac{1}{\sqrt{\lambda}}\right] \, .
 \label{Li7}
\ee
%in an agreement with eq. (\ref{Li1}).

For the anomalous dimension at $j-2 \sim 1/\sqrt{\lambda}$ from (\ref{Li5})
we obtain the square root singularity similar to that appearing at small
$j-1 =\omega _0$ (\ref{7i})
\be
\gamma ~=~ -\frac{\lambda^{1/4}}{\sqrt{2}}
\left(1+ \frac{a_{01}}{\sqrt{\lambda}}\right)
\Bigl(\sqrt{D_1+j-2} - \sqrt{D_1}\Bigr)\,,
 \label{Li8}
\ee
where $D_1$ (\ref{Li1}) is equal to
%closed to
the correction $\Delta$ to the graviton trajectory intercept with our accuracy
%{\bf ?? in diffusion approximation ??}:
$$\Delta = D_1 \approx  \frac{2}{\sqrt{\lambda}}
\left(1+ \frac{1}{2\sqrt{\lambda}}\right).$$
Note, that in the region $j-2 <- \Delta$, the anomalous dimesnion is complex
similar to it in the perturbative regime at $j-1<\omega_0$ (\ref{7i}).
Moreover, the position of the BFKL singularity of $\gamma$ at large coupling constants
can be found from the calculation of the radius of the divergency of the perturbation theory
in $1/\sqrt{\lambda}$ at small $j-2$.

\section{Numerical analysis of the Pomeron intercept $j_0(\lambda)$
}
\indent

Let us obtain  an unified expression for the position of the Pomeron
%$j$-plane
singularity  $j_0=1+\omega_0$
for arbitrary values of $\lambda$,
using an interpolation between weak and strong coupling regimes.

It is convenient to replace $\omega_0$ with the new variable $t$ as follows
\be
t_0 ~=~ \frac{\omega_0}{1-\omega_0},~~~~
\omega_0 ~=~ \frac{t_0}{1+t_0}
\, . \label{1.num}
\ee

This variable  has the asymptotic behavior $t_0 \sim \lambda$ at $\lambda \to 0$ and
$t_0 \sim \sqrt{\lambda}/2$ at $\lambda \to \infty$ similar
to the case of the cusp anomalous dimension
(see, for example, \cite{Kotikov:2003fb}). So, following
the method of Refs. \cite{Kotikov:2003fb,Kotikov:2004er,Bern:2006ew}, we
shall write a simple algebraic equation
for $t_0=t_0(\lambda)$ whose
solution will interpolate $\omega_0$
%should approximate $t$ behaviour
for the full $\lambda$ range.

We choose the equation of the form
\be
k_0(\lambda)~=~k_1(\lambda) t_0 + k_2(\lambda) t_0^2
 \, , \label{2.num}
\ee
where the following anzatz for the coefficinets $k_0$, $k_1$ and
$k_2$ is used:
\be
k_0(\lambda) ~=~ \beta_0 \lambda + \alpha_0 \lambda^2,~~
k_1(\lambda) ~=~ \beta_1+ \alpha_1 \lambda,~~
k_2(\lambda) ~=~ \gamma_2 \lambda^{-1} + 
\beta_2 + \beta_2 \lambda \, . \label{3.num}
\ee
Here  $\gamma_2$, $\alpha_i$ and $\beta_i$ $(i=0,1,2)$ are
%$\alpha_0$, $\alpha_1$, $\beta_0$, $\beta_1$ and $\beta_2$ are
free parameters,
which are fixed using the known asymptotics of  $\omega_0$
%and/or $t$
at $\lambda \to 0$ and $\lambda \to \infty$.

The solution of quadratic equation (\ref{2.num}) is given below
%simple
\be
%t_0 ~=~ \frac{\sqrt{k_1^2+4k_0k_2}-k_1}{2k_2}.~~
t_0 ~=~ \frac{k_1}{2k_2} \, \left[
\sqrt{1+\frac{4k_0k_2}{k_1^2}}-1 \right] \, .~~
\label{3a.num}
\ee

To fix the parameters $\gamma_2$, $\alpha_i$ and $\beta_i$ $(i=0,1,2)$, we
use two known coefficients for the weak coupling expansion of  $\omega_0$:
%at $\lambda \to 0$
\be
\omega_0 ~=~ \tilde{e}_1\lambda + \tilde{e}_2\lambda^2  + \tilde{e}_3\lambda^3
+ \dots \,
~~~(\mbox{at}~~ \lambda \to 0)\,  \label{4.num}
\ee
with
\be
 \tilde{e}_1 ~=~ \frac{\ln2}{\pi^2} \approx 0.07023,~~
\tilde{e}_2 ~=~ - \tilde{e}_1 \,  \frac{7.5812}{16\pi^2} \approx -0.00337
\label{5.num}
\ee
and first four terms of its strong coupling expansion
%at $\lambda \to \infty$
\be
\omega_0 ~=~ 1-\Delta,~~ \Delta ~=~ \frac{2}{\sqrt{\lambda}}\left(
1+\frac{\tilde{t}_1}{\sqrt{\lambda}} +\frac{\tilde{t}_2}{\lambda}
+\frac{\tilde{t}_3}{\lambda^{3/2}} +\frac{\tilde{t}_4}{\lambda^2}
+ \dots \right) \,
~~~(\mbox{at}~~ \lambda \to \infty)\,
%\, ,
\label{7.num}
\ee
with (see below Eq. (\ref{11.de})
\be
 \tilde{t}_1 ~=~ \frac{1}{2},~~ \tilde{t}_2 ~=~ -\frac{1}{8},~~
\tilde{t}_3 ~=~ - 1 -3 \zeta_3,~~ \tilde{t}_4 ~=~ 2a_{12} - \frac{145}{128} 
- \frac{9}{2}\zeta_3 \, .
%8\ln2 -3 \approx 2.54518 \, .
\label{8.num}
\ee
The coefficients $\tilde{e}_3$ and $\tilde{t}_4$ are unknown but
%are estimated below.
we estimate them later from the interpolation.

\begin{figure}[!t] %\label{HTC_comp}
\unitlength=1mm
\vskip -1.5cm
\begin{picture}(0,100)
% \put(0,-5){\epsfig{file=PomerG2.eps,width=150mm,height=95mm}}
\put(0,-5){\epsfig{file=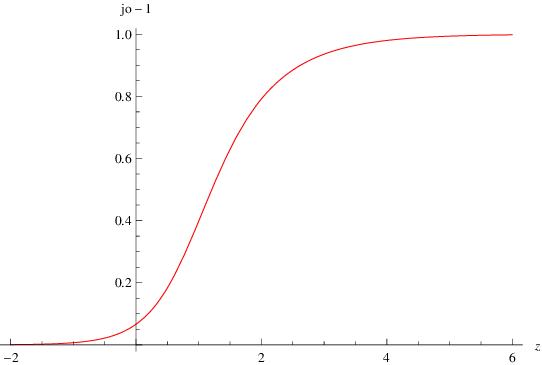,width=150mm,height=95mm}}
\end{picture}
\vskip 0.2cm
\caption{(color-online). The results for $j_0$ as a function of $z$ 
($\lambda=10^z$).
}
\end{figure}

Then, for the weak and strong coupling
%corresponding
expansions of $t$ one obtains
\bea
t_0 =& e_1\lambda + e_2\lambda^2 + e_3\lambda^3 + \dots ,~~~~~~~
&(\mbox{when} ~~\lambda \to 0) \, ,
\label{9.num} \\
t_0 =& \frac{\sqrt{\lambda}}{2} \, \left(
1-\frac{t_1}{\sqrt{\lambda}}-\frac{t_2}{\lambda}
-\frac{t_3}{\lambda^{3/2}}-\frac{t_4}{\lambda^2}\right)
+ \dots ,
~~~~&(\mbox{when} ~~\lambda \to \infty) \, ,
\label{10.num}
\eea
where
\bea
&&e_1 ~=~ \tilde{e}_1,
%\approx 0.25,
~~ e_2 ~=~ \tilde{e}_2 + \tilde{e}_1^2,~~ e_3 ~=~ \tilde{e}_3 + \tilde{e}_1\tilde{e}_2 + \tilde{e}_1^3,~~ 
%+ \tilde{e}_1^2 \approx 0.05,
t_1 ~=~ \tilde{t}_1 + 2 = \frac{5}{2},~~  \nonumber \\
&& t_2 ~=~  \tilde{t}_2-\tilde{t}_1^2 =  -\frac{3}{8},~~
%\nonumber \\&&
t_3~=~  \tilde{t}_3-2\tilde{t}_2\tilde{t}_1+\tilde{t}_1^3 = - \frac{3}{4} \Bigl( 1 +4 \zeta_3\Bigr), \nonumber \\
&&t_4~=~  %\Bigl(
\tilde{t}_4 -2\tilde{t}_3\tilde{t}_1-\tilde{t}_2^2 
+3\tilde{t}_2\tilde{t}_1^2-\tilde{t}_1^4
%\Bigr) - \Bigl(\tilde{t}_2 -2\tilde{t}_1 +1\Bigr) 
= 2a_{12} - \frac{39}{128} 
- \frac{3}{2}\zeta_3 \, .
\label{10a.num}
\eea

Comparing the l.h.s. and the r.h.s. of Eq. (\ref{2.num}) at
$\lambda \to 0$ and $\lambda \to \infty$, respectively, we derive the
following relations
\be
\alpha_2~=~4 \alpha_0,~~ \alpha_2~=~10 \alpha_0,~~
\beta_1~=~C_1 \alpha_0,~~~ \beta_2~=~C_2 \alpha_0,~~ \gamma_2~=~C_3 \alpha_0,
\beta_0~=~\Bigl(C_2-22\Bigr) %\alpha_0/4 
\frac{\alpha_0}{4} \, 
\label{11.num}
\ee
with the free parameter $\alpha_0$ which disappears 
in the retionship $k_1/k_2$ and $k_0/k_2$ and, thus, in the results
(\ref{3a.num}) for $t_0$.

Here
\be
C_1 \approx 88.60,~~ C_2 \approx 42.41,~~ C_3 \approx -277.0 \, ,
\label{C.num}
\ee
which lead to the following predictions for
the 
%third 
coefficients $e_3$ and $t_4$ in (\ref{9.num}) and (\ref{10.num})
%In this example,
\be
e_3 ~=~ - \frac{10e_2+2C_2e_1e_2+4e_1^2
%+C_3e_2^2
}{C_1+2C_3e_1} \approx -0.00079,~~
%8e_2(2e_1^2+t_1e_2)}{1-8t_1e_1} \approx -0.0082686,~~~
t_4 ~=~ \frac{9+16 (C_3-5C_1+7C_2)}{128}  \approx - 40.5774
%-\frac{t_1^2}{2} - \frac{e_1(1-8t_1e_1)}{2e_2} \approx -25.7867 \,
%,
\label{15a.num}
\ee
and,
%or,
respectively, for the corresponding terms in (\ref{4.num}), (\ref{7.num}) 
and (\ref{8.num})
\be
%e_3 ~\approx~ ???,~~
\tilde{e}_3 ~\approx~ -0.00066,~~~~
%t_2 ~\approx~ ???,~~
\tilde{t}_4 ~\approx~ -51.0117,~~~~ a_{12} ~\approx~ -22.2348 \, .
\label{16.num}
\ee

Note that the results for the coefficients $e_3$, $t_4$, $\tilde{e}_3$,
$\tilde{t}_4$ and $a_{12}$  do not depend on the free parameter $\alpha_0$.

%\vskip 0.5cm

%{\bf 3.}~~ On Figs. 1 and 2,
On Fig. 1, we plot the pomeron intercept $j_0$
as a function of the coupling constant $\lambda$.
The behavior of the pomeron intercept $j_0$ shown in Fig.1 is
similar to that found in QCD with some additional assumptions (see ref. \cite{Stasto:2007uv}).

\section{ Conclusion}
\indent

We found the intercept of the BFKL pomeron at
weak and strong coupling regimes in
the ${\mathcal N}=4$ Super-symmetric Yang-Mills model.

At large couplings $\lambda \rightarrow \infty$, the correction $\Delta$ for the
Pomeron
intercept $j_0=2-\Delta$ has the form (see Appendix (\ref{11.dc3}))
\be
\Delta ~=~  \frac{2}{\lambda^{1/2}} \, \left[1
+
%\Bigl(8\ln 2 -3\Bigr) \,
\frac{1}{2\lambda^{1/2}} - \frac{1}{8\lambda} -
\Bigl(1 +3 \zeta_3 \Bigr) \frac{1}{\lambda^{3/2}} +
%\left(1-%\frac{33}{4}-36\ln 2 + 48 \ln^2 2 - 2 a_{02} \right)
\left(2a_{12} - \frac{145}{128} - \frac{9}{2}\zeta_3 \right)
\frac{1}{\lambda^2} + O\left(\frac{1}{\lambda^{5/2}}\right)
\right] \, . \label{11.de}
\ee
The anomalous dimension has a square-root singularity at the value of the BFKL intercept both
in the weak and strong coupling regimes. This value is related to the radius of convergency
of perturbation theory in $\lambda$ and $1/\sqrt{\lambda}$ near the points 
$j_0=1$ and $j_0=2$, respectively.

The fourth corrections in (\ref{11.de})
contain unknown coefficient $a_{12}$,
which will be obtained after
%needs 
the evaluation of spinning folded string on the two-loop level.
Some estimations were given in Section 6.

The
%corresponding large coupling expansion for the
slope of the universal anomalous dimension at $j=2$ 
known by the direct calculations \cite{VelizhComm} up to the fifth order of perturbation theory
can be
%close form
written as follows
\be
\gamma'(2) ~=~  - \frac{\sqrt{\lambda}}{4} \,
\frac{I_3(\sqrt{\lambda})}{I_2(\sqrt{\lambda})}
\, , \label{11a.de}
\ee
according to the well known Basso result \cite{Basso:2011rs} for
local operators of an arbitrary twist.

\vspace{5mm} {\large \textbf{Acknowledgments.}}\\
%[5mm]
This work was supported in part
by the Alexander von Humboldt fellowship and
by RFBR grant No. 13-02-01060-a
%10-02-01454-a 
(A.V.K.).
We are grateful to Arkady Tseytlin and Dmitro Volin for useful discussions.
%correspondences.

%%%%%%%%%%%%%%%%%%%%%%%%%%%%%%%%%%%%%%%%%%%%%%%%%%%%%%%%%%%%%%%
%\appendix
%%%%%%%%%%%%%%%%%%%%%%%%%%%%%%%%%%%%%%%%%%%%%%%%%%%%%%%%%%%%%%%
%\section{Inhomogeneities}
%\label{App:Inhomogeneities}
%%%%%%%%%%%%%%%%%%%%%%%%%%%%%%%%%%%%%%%%%%%%%%%%%%%%%%%%%%%%%%%

%\section{Appendix A}
%\label{App:A}
%\def\theequation{A\arabic{equation}}
%\setcounter{equation}{0}

\setcounter{secnumdepth}{2}
\addcontentsline{toc}{section}{APPENDIX}

\setcounter{section}{0}
\setcounter{subsection}{0}
\setcounter{equation}{0}
 \def\thesection{\Alph{section}}
\def\thesubsection {\thesection.\arabic{subsection}}
\def\theequation{\thesection.\arabic{equation}}

%\appendix
\section{Appendix}

Here we discuss
%give the results for
coefficients $D_m$ and the Pomeron
intercept $2-\Delta$ using expression (\ref{4a}) at comparatively large $j$ in the region $j<<\sqrt{\lambda}$.
%values

\subsection{String energy at $1<<j<<\sqrt{\lambda}$
%Equations for coefficients $D_m$ and the Pomeron intercept $2-\Delta$
%Small spin expansion
}

The recent results for the string energies
\cite{Gromov:2011bz}
in the region restricted by inequalities (\ref{jll1})
can be presented  in the form
\footnote{Here we put $S=j-2$, which in particular is related to the
use of the angular momentum $J_{an}=2$ in calculations of Refs \cite{Gromov:2011de,Roiban:2011fe}.}
\be
\frac{E^2}{4} ~=~ \sqrt{\lambda} \, \frac{S}{2}\, \left[h_0(\lambda)
+ h_1(\lambda) \frac{S}{\sqrt{\lambda}} + h_2(\lambda) \frac{S^2}{\lambda}
\right] + O\Bigl(S^{7/2}\Bigr),
\label{5a}
\ee
where
\be
 h_i(\lambda) ~=~  a_{i0} + \frac{a_{i1}}{\sqrt{\lambda}} +
\frac{a_{i2}}{\lambda} +  \frac{a_{i3}}{\sqrt{\lambda^3}} +
\frac{a_{i2}}{\lambda^2}.
\label{5.1a}
\ee

 The contribution $\sim \sqrt{S}$ can be extracted directly from
the
% famous
Basso result \cite{Basso:2011rs}
taking $J_{an}=2$ according to \cite{Gromov:2011bz}:
\be
h_0(\lambda) =
\frac{I_3(\sqrt{\lambda})}{I_2(\sqrt{\lambda})} + \frac{2}{\sqrt{\lambda}} =
\frac{I_1(\sqrt{\lambda})}{I_2(\sqrt{\lambda})} - \frac{2}{\sqrt{\lambda}}\, ,
\label{Ad5.1}
\ee
where $I_k(\sqrt{\lambda})$ is the modified Bessel functions.
It leads to the following values of  coefficients $a_{0i}$
\be
a_{00} ~=~ 1,~~ a_{01}~=~ - \frac{1}{2},~~a_{02} ~=~ a_{03}~=~  \frac{15}{8},~~
a_{04}~=~  \frac{135}{128}
\label{Ad5.2}
\ee

The coefficients $a_{10}$ and $a_{20}$ come from considerations of the
classical part of
%For
the folded spinning string corresponding to the twist-two
operators \footnote{  We are grateful to Arkady Tseitlin for
explaining this point.}
(see, for example, \cite{Roiban:2011fe})
\be
a_{10}~=~  \frac{3}{4},~~a_{20} ~=~ - \frac{3}{16}\,.
\label{Ad5.3}
\ee

The one-loop coefficient $a_{11}$ is found recently in
the paper
 \cite{Gromov:2011bz} (see also \cite{Beccaria:2012tu}), considering different
asymptotical regimes with taking into account the Basso result \cite{Basso:2011rs}
\be
a_{11}~=~ \frac{3}{16} \Bigl(1-\zeta_3\Bigr),
\label{Ad5.4}
\ee
where $\zeta_3$ is the Euler $\zeta$-function.

All
%above
calculations
%for the one-loop corrections $a_{01}$
were performed
for nonzero values of the angular momentum $J_{an}$ (really, $J_{an}=2$
was used)
 and
are applicable  also to the finite $S$ values.
\footnote{ The previous calculations \cite{Tirziu:2008fk} were done with
the zero values of the angular momentum $J_{an}$ and cannot be directly
applied for the finite $S$ values. We are grateful to Arkady Tseitlin for
explaining this point to us.}
Moreover, all these coefficients are in a full agreement with numerical
$Y$-system predictions (see \cite{Gromov:2009zb,Gromov:2011de} and
references therein).

\subsection{ Equations for coefficients $D_m$ and the Pomeron intercept $2-\Delta$
%Small spin expansion
}

Thus, from expression (\ref{5a}) we obtain the following expansions of
even powers of $E$ in the small parameter
$j/\sqrt{\lambda}$
\be
{\left(\frac{E^2}{4}\right)}^2 = \lambda  \, \frac{S^2}{4} \, \left[
h_0^2(\lambda)
+ 2h_0h_1(\lambda) \frac{S}{\sqrt{\lambda}}
%+ e_2(\lambda) \frac{j^2}{\lambda}
\right],~~~
%\nonumber \\
{\left(\frac{E^2}{4}\right)}^3 ~=~ \lambda^{3/2}  \, \frac{S^3}{8} \,
h_0^3(\lambda) \, .
\label{7a}
\ee

Comparing the coefficients in the front of $S$, $S^2$ and $S^3$ in
the l.h.s. and
r.h.s of (\ref{4a}), we derive the equations
\bea
1 &=& \frac{\sqrt{\lambda}}{2} \, h_0 \, \overline{D}_1,~~~
\overline{D}_1 ~=~ \left(D_1+2D_2+3D_3\right), \label{8a} \\
0 &=& \frac{1}{2} \, h_1 \, \overline{D}_1 +
\frac{\lambda}{4} \, h_0^2 \, \overline{D}_2,~~~
\overline{D}_2 ~=~ \left(D_2+3D_3\right), \label{8b} \\
0 &=& \frac{1}{2\sqrt{\lambda}} \, h_2 \, \overline{D}_1 +
\frac{\sqrt{\lambda}}{4} \, h_0h_1 \, \overline{D}_2
+ \frac{\lambda^{3/2}}{8} \, h_0^3 \, D_3.
\label{9a}
\eea

Their perturbative solution leads is given below
\bea
\overline{D}_1 &=& \frac{2}{\sqrt{\lambda}} \, \frac{1}{h_0}
%~=~ \frac{2}{\sqrt{\lambda}} \, \frac{1}{h_0^2},
~~~
%\label{10a} \\
\overline{D}_2 ~=~ -\frac{2}{\lambda} \, \frac{h_1}{h_0^2} \,
\overline{D}_1 ~=~ -\frac{4}{\lambda^{3/2}} \, \frac{h_1}{h_0^3},
\label{10b} \\
D_3 &=&
\frac{4}{\lambda^2} \, \frac{2h_1^2-h_0h_2}{h_0^4} \, \overline{D}_1
~=~ \frac{8}{\lambda^{5/2}} \, \frac{2h_1^2-h_2h_0}{h_0^5}.
\label{11a}
\eea
and, correspondingly,
\be
D_2 ~=~ \overline{D}_2 - 3 D_3 , ~~
%\nonumber \\
D_1
%&=& \overline{D}_1 -2D_2-3D_3
~=~ \overline{D}_1 -2 \overline{D}_2 + 3 D_3 \, .
\label{11b}
\ee

Finally, we obtain the correction $\Delta$ to the Pomeron intercept in the form
\bea
\Delta &=& D_1+D_2+D_3 ~=~ \overline{D}_1- \overline{D}_2 + D_3 \nonumber \\
&=&
\frac{2}{\sqrt{\lambda}} \, \frac{1}{h_0^2} +
\frac{4}{\lambda^{3/2}} \, \frac{h_1}{h_0^3} +
\frac{8}{\lambda^{5/2}} \, \frac{2h_1^2-h_2h_0}{h_0^5},
\label{11.1a}
\eea
where the $\lambda$-dependence of parameters $h_i$ is given in Eqs. (\ref{5.1a}) and
(\ref{Ad5.1}).

\subsection{Strong coupling expansions  of  $D_m$ and $\Delta$
%for Pomeron intercept
}

Using expressions (\ref{Ad5.2})-(\ref{Ad5.4}) we have
\bea
D_3 &=& \frac{8r_3}{\lambda^{5/2}}
%\frac{21}{2\lambda^{5/2}}
+ O\left(\frac{1}{\lambda^{7/2}}\right)\,
, ~~~
%\label{11.da} \\
\overline{D}_2 ~=~
%&=&
- \frac{4}{\lambda^{3/2}} \, \left[
\overline{c}_2 + \frac{\overline{c}_3}{\lambda^{1/2}} +
\frac{\overline{c}_4}{\lambda} + O\left(\frac{1}{\lambda^{3/2}}\right)
\right] \, , \label{11.da1} \\
\overline{D}_1 &=&  \frac{2}{\lambda^{1/2}} \, \left[1
+ \frac{\overline{d}_1}{\lambda^{1/2}} +
\frac{\overline{d}_2}{\lambda} + \frac{\overline{d}_3}{\lambda^{3/2}} +
\frac{\overline{d}_4}{\lambda^2} + O\left(\frac{1}{\lambda^{5/2}}\right)
\right] \, , \label{11.da2}
\eea
where
\bea
\overline{c}_2 &=& a_{10} = \frac{3}{4}, ~~~
\overline{c}_3 ~=~ a_{11}-3a_{10}a_{01} =
\frac{3}{16} \Bigl(7-8\zeta_3\Bigr), ~~~
r_3 ~=~  2a^2_{10}-a_{20}= \frac{21}{16},
%~=~-\frac{17}{4} + 9\ln 2+\frac{3}{4}\zeta_3,
\nonumber \\
\overline{c}_4 &=& a_{12} +3a_{10}\Bigl(2a^2_{01}-a_{02}\Bigr) -3a_{11}a_{01}
= a_{12} - \frac{9}{16} \Bigl(5+4\zeta_3\Bigr)
\label{11.db1}
\eea
and
\bea
\overline{d}_1 &=& -2 a_{01} = \frac{1}{2}, ~~~
\overline{d}_2 ~=~ 2a^2_{01}-a_{02} = -\frac{13}{8} , ~~~
\overline{d}_3 ~=~
%&=&
2a_{01}a_{02}-a^3_{01}- a_{03} = -\frac{29}{8}
, \nonumber \\
\overline{d}_4 &=& a^4_{01}-3a^2_{01}a_{02}+2a_{01}a_{03}+a^2_{02}- a_{04}
= - \frac{97}{128} \, .
\label{11.db2}
\eea

Here $a_{02},a_{12}, a_{03}$ and $a_{04}$ are parameters which should be calculated
in future at two, three and four loops of the string perturbation theory. It is
important, that the
coefficients $D_k$ tend to zero at large $\lambda$ as $\lambda^{-n+1/2}$

Analogously, we can obtain expressions for $D_2$, $D_1$ and $\Delta$:
\bea
D_2 &=& - \frac{4}{\lambda^{3/2}} \, \left[
c_2 + \frac{c_3}{\lambda^{1/2}} +
\frac{c_4}{\lambda} + O\left(\frac{1}{\lambda^{3/2}}\right)
\right] \, , \label{11.dc1} \\
D_1 &=&  \frac{2}{\lambda^{1/2}} \, \left[1
+ \frac{d_1}{\lambda^{1/2}} +
\frac{d_2}{\lambda} + \frac{d_3}{\lambda^{3/2}} +
\frac{d_4}{\lambda^2} + O\left(\frac{1}{\lambda^{5/2}}\right)
\right] \, , \label{11.dc2}\\
\Delta &=&  \frac{2}{\lambda^{1/2}} \, \left[1
+ \frac{\hat{d}_1}{\lambda^{1/2}} +
\frac{\hat{d}_2}{\lambda} + \frac{\hat{d}_3}{\lambda^{3/2}} +
\frac{\hat{d}_4}{\lambda^2} + O\left(\frac{1}{\lambda^{5/2}}\right)
\right] \, , \label{11.dc3}
\eea
where
\bea
c_2 &=& \overline{c}_2,~~ c_3 ~=~ \overline{c}_3,~~
c_4 ~=~ \overline{c}_4 + 6r_3
%\frac{63}{16}
,~~d_1 ~=~ \overline{d}_1 ~=~
\hat{d}_1 \, , \label{11.dd1} \\
d_2 &=& \overline{d}_2 + 4 \overline{c}_2,~~
d_3 ~=~ \overline{d}_3 + 4\overline{c}_3,~~
d_4 ~=~ \overline{d}_4 + 4\overline{c}_4 + 12 r_3
%\frac{63}{4}
\, , \label{11.dd2} \\
\hat{d}_2 &=& \overline{d}_2 + 2 \overline{c}_2,~~
\hat{d}_3 ~=~ \overline{d}_3 + 2 \overline{c}_3,~~
\hat{d}_4 ~=~ \overline{d}_4 + 2 \overline{c}_4 + 4r_3
%\frac{21}{4}
%\, .
\label{11.dd3}
\eea
and all $\overline{c}_i$ and $\overline{d}_i$ are given above in eqs
(\ref{11.db1}) and (\ref{11.db2}).
So,  we have
\be
\hat{d}_1 ~=~ \frac{1}{2},~~\hat{d}_2 ~=~ -\frac{1}{8},
~~\hat{d}_3 ~=~ -1 -3\zeta_3, ~~\hat{d}_4 ~=~ 2a_{12} -
\frac{145}{128} - \frac{9}{2} \zeta_3 \, .
\label{New}
\ee

Using a similar approach, the coefficients $\hat{d}_1$ and $\hat{d}_2$
were found recently in the paper \cite{Costa:2012cb}.
The corresponding coefficients $c_{2,0}$ and $c_{3,0}$ in \cite{Costa:2012cb}
coincide with our $\hat{d}_1$ and $\hat{d}_2$ but in the expression for the
Pomeron intercept they contributed with an opposite sign.
Further, in the talk
of Miguel S. Costa ``Conformal Regge Theory'' on IFT Workshop
``Scattering Amplitudes in the Multi-Regge limit''
 (Universidad Autonoma de Madrid, 24 - 26  Oct 2012)
(see http://www.ift.uam.es/en/node/3985)
%a correct 
the sign of these contributions 
%of these terms 
to the Pomeron intercept was correct but there is a misprint the definition of
%given but 
the parameter of
%corresponding
expansion.
Note, however, that we have the next term $\hat{d}_3$ in the strong coupling expansion.

\subsection{Anomalous dimension near $j=2$}

At $j=2$, the universal anomalous dimension is zero (\ref{1e}), but its
derivative $\gamma'(2)$ (the slope of $\gamma$) has a nonzero value
in the perturbative theory
\be
\gamma'(2) = - 
%\frac{1}{3} 
\frac{\lambda}{24} + \frac{1}{2}
{\left(\frac{\lambda}{24}\right)}^2 -  \frac{2}{5}
{\left(\frac{\lambda}{24}\right)}^3 + \frac{7}{20}
{\left(\frac{\lambda}{24}\right)}^4 -  \frac{11}{35}
{\left(\frac{\lambda}{24}\right)}^5 +
O\bigl(\lambda^6)\,,
%0\left({\left(\frac{\lambda}{8}\right)}^4\right),
\label{Ad1}
\end{equation}
as it follows from exact three-loop calculations \cite{Kotikov:2004er,Kotikov:2006}.
Two last terms were calculated by V. Velizhanin \cite{VelizhComm} from the explicit
results for $\gamma$ in five loops \cite{Lukowski:2009ce}. 

To find the slope
%results for
$\gamma'(2)$ at large values of the coupling constant
we calculate
%should compare
the derivatives of the l.h.s. and r.h.s. of eq. (\ref{7e})
%combined with (\ref{8e}), i.e.
written in the form
\be
j-2 = \sum_{m=1}  D_m \left[{\left(\frac{j}{2}-\gamma\right)}^{2m}-1\right]\,
\label{Ad2}
\ee
in the variable $j$ for $j=2$ using $\gamma(2)=0$:
\be
1 =  \Big(1-2\gamma'(2)\Bigr) \, \sum_{m=1} m D_m  \equiv
 \Big(1-2\gamma'(2)\Bigr) \, \overline{D}_1 \,,
\label{Ad3}
\ee
where $\overline{D}_1$ is found in (\ref{8a}).
%in Appendix A in the expansion up to $m=3$ (see (\ref{8a})).
So we obtain explicitly
%in the closed form
\be
1-2\gamma'(2) = \frac{\sqrt{\lambda}}{2} \, h_0(\lambda) \,.
\label{Ad4}
\ee
Substituting (\ref{Ad5.1}) in (\ref{Ad4}), we have the closed  form
for the slope $\gamma'(2)$
\[
\gamma'(2) = - \frac{\sqrt{\lambda}}{4}
\frac{I_3(\sqrt{\lambda})}{I_2(\sqrt{\lambda})} \, ,
%\label{Ad5.2a}
\]
which is in full agreement with predictions (\ref{Ad1}) of
perturbation theory.

\section{Appendix
%B.
%Anomalous dimension of the Konishi operator
}
\label{App:B}
\def\theequation{B\arabic{equation}}
\setcounter{equation}{0}

We apply Eqs. (\ref{7e}) and (\ref{8e}) with $j=4$ (and/or $S=2$)
and $D_i$ ($i=1,2,3$)
obtained in Appendix A,
%the previous section,
to find the large $\lambda$ asymptotics
of the anomalous dimension of the Konishi operator. So, it obeys to the
equation
\be
2 ~=~ \sum_{m=1}  D_m \left(x^m-1\right),~~~x\equiv (2-\gamma_k)^2
\label{1K}
\ee

\vskip 0.5cm

{\bf 1.} It is convenient to consider firstly the particular case, when
$\overline{D}_2=\overline{D}_3=0$ and, thus, $D_1=\overline{D}_1=
2/{\sqrt{\lambda}h_0}$. So, we have
\be
2 ~=~ \overline{D}_1 (x-1)
\label{2K}
\ee
and
\be
x ~=~ \frac{2}{\overline{D}_1}+1 ~=~ \sqrt{\lambda}h_0 +1 \, ,
\label{3K}
\ee
where $h_0$ has the closed form (\ref{Ad5.1}). So, the
%The
anomalous dimension $\gamma_K$ can be represented as
%has the form
\be
2-\gamma_K  ~=~ \Bigl(\sqrt{\lambda}h_0 +1\Bigr)^{1/2}
%\lambda^{1/4}\sqrt{h_0} \sqrt{1+\frac{1}{\sqrt{\lambda}h_0}}
~\approx ~  \lambda^{1/4} \left(\sqrt{h_0} +
\frac{1}{2\sqrt{\lambda}\sqrt{h_0}} -
\frac{1}{8\lambda h_0^{3/2}} + O\left(\frac{1}{\lambda^{2}} \right) \right) \, .
\label{4K}
\ee

For the case of the classic string, where $h_0=1$, i.e. $a_{00}=1$ and
$a_{0i}=0$ ($i\geq1$), we reconstruct well-known results
\footnote{We should remind that our anomalous dimension $\gamma_K$ has the
additional factor $-1/2$, i.e. $\gamma_K=-\gamma^{standart}_K/2$.}
\be
2-\gamma_K
~\approx ~  \lambda^{1/4} \left( 1 + \frac{1}{2\sqrt{\lambda}} -
\frac{1}{8\lambda} + O\left(\frac{1}{\lambda^{3/2}} \right) \right) \, .
\label{5K}
\ee

For the exact values of $h_0$ done in Eqs. (\ref{5.1a}) and (\ref{Ad5.2}),
we have
\bea
2-\gamma_K
&\approx &  \lambda^{1/4} \left( 1 + \frac{1+a_{01}}{2\sqrt{\lambda}} +
%\left[\frac{1}{2} +a_{01} \right] +
\frac{1}{2\lambda} \left[a_{02}-\frac{(1+a_{01})^2}{4} \right]
+ O\left(\frac{1}{\lambda^{3/2}} \right) \right)
\nonumber \\
&=&  \lambda^{1/4} \left( 1 + \frac{1}{4\sqrt{\lambda}}
+ \frac{29}{32\lambda} + O\left(\frac{1}{\lambda^{3/2}} \right) \right) \, .
\label{6K}
\eea

\vskip 0.5cm

{\bf 2.} In the case when all $\overline{D}_i$ $(i=1,2,3)$ are nonzero,
it is convenient to represent the solution of the equation (\ref{1K}) in the
following form
\be
x ~=~ \sqrt{\lambda}h_0 +1 + x_1 + \frac{x_2}{\sqrt{\lambda}} \, .
\label{7K}
\ee

Expanding $\overline{D}_i$ in the inverse series of $\sqrt{\lambda}$ and
compare the coefficients in the front of $\lambda^0$ and $1/\sqrt{\lambda}$,
we have
\be
x_1 ~=~ 2a_{10},~~ x_2 ~=~ 2a_{11}+4a_{20}
%4 \Bigl[a^2_{10}+2a_{20}+a_{10}a_{01}+a_{11}\Bigr]\equiv 4c_{20}
\, .
\label{8K}
\ee

So, the solution of the equation (\ref{7K}) with the coefficients (\ref{8K})
has the form
\be
2-\gamma_K
~\approx ~
\lambda^{1/4} \Biggl( 1 + \frac{a_{01} +1 +2a_{10}}{2\sqrt{\lambda}}
%\Bigl[a_{01} +1 +2a_{10} \Bigr]
 +\frac{1}{2\lambda}
\biggl[a_{02} + 2a_{11}+4a_{20} - \frac{(1+a_{01}+2a_{10})^2}{4}
\biggr]
+ O\left(\frac{1}{\lambda^{3/2}} \right) \Biggr) \,.
\label{9K}
\ee

Using
%$a_{01}^{\Sigma}=0$ and $a_{10}^{\Sigma}=1/4$ (see
Eq.s (\ref{Ad5.2})-(\ref{Ad5.4})
the exact values of $a_{ij}$,
we have
\be
2- \gamma_K  ~\approx ~   \lambda^{1/4} \left( 1 + \frac{1}{\sqrt{\lambda}}
%\left[\frac{11}{4}-4\ln 2\right]
+ \frac{1}{4\lambda} \Bigl[1-6\zeta_3
\Bigr]
+ O\left(\frac{1}{\lambda^{3/2}} \right) \right)
\,
\label{10K}
\ee

We would like to note that our coefficient in the front of $\lambda^{-1/4}$
is equal to $1$, which in an agreement with calculations performed in
\cite{Gromov:2009zb,Gromov:2011de,Roiban:2011fe}. Further,
the coefficient in  front of $\lambda^{-3/4}$ agrees with
the results of \cite{Gromov:2011bz} (see also Refs. \cite{Beccaria:2012tu}
and \cite{Frolov:2012zv}).


\begin{thebibliography}{99}


%\cite{Chew:1961ev}
\bibitem{Chew:1961ev}
  G.~F.~Chew and S.~C.~Frautschi,
  %``PRINCIPLE OF EQUIVALENCE FOR ALL STRONGLY INTERACTING PARTICLES WITHIN THE
  %S MATRIX FRAMEWORK,''
  Phys.\ Rev.\ Lett.\  {\bf 7} (1961) 394;
  %%CITATION = PRLTA,7,394;%%
%\cite{Gribov:1961fr}
%\bibitem{Gribov:1961fr}
  V.~N.~Gribov,
  %``Partial Waves With Complex Orbital Angular Momenta And The Asymptotic
  %Behavior Of The Scattering Amplitude,''
  Sov.\ Phys.\ JETP {\bf 14} (1962) 1395
  [Zh.\ Eksp.\ Teor.\ Fiz.\  {\bf 41} (1961) 1962];
  %%CITATION = ZETFA,41,1962;%%
%\cite{Gribov:1961fm}
%\bibitem{Gribov:1961fm}
%  V.~N.~Gribov,
  %``Asymptotic Behavior Of The Scattering Amplitude At High-Energies,''
  Nucl.\ Phys.\  {\bf 22} (1961) 249.
  %%CITATION = NUPHA,22,249;%%


\bibitem{Pomeranchuk}
I. Ya. Pomeranchuk,  Sov.\ Phys.\ JETP {\bf 7} (1958) 499
[Zh.\ Eksp.\ Teor.\ Fiz.\  {\bf 34} (1958) 725];
L. B. Okun and I. Ya. Pomeranchuk,  Sov.\ Phys.\ JETP {\bf 3} (1956) 307
[Zh.\ Eksp.\ Teor.\ Fiz.\  {\bf 30} (1956) 424].


\bibitem{BFKL}
L.~N.~Lipatov, Sov.\ J.\ Nucl.\ Phys.\ \textbf{23} (1976) 338;
%%CITATION = YAFIA,23,642;%%
V.~S.~Fadin, E.~A.~Kuraev and L.~N.~Lipatov,
Phys.\ Lett.\ B \textbf{60} (1975) 50;
%%CITATION = PHLTA,B60,50;%%
E.~A.~Kuraev, L.~N.~Lipatov and V.~S.~Fadin,
Sov.\ Phys.\ JETP \textbf{44} (1976) 443;
%%CITATION = ZETFA,71,840;%%
E.~A.~Kuraev, L.~N.~Lipatov and V.~S.~Fadin,
Sov.\ Phys.\ JETP \textbf{45} (1977) 199;
%%CITATION = ZETFA,72,377;%%
I.~I.~Balitsky and L.~N.~Lipatov,
Sov.\ J.\ Nucl.\ Phys.\ \textbf{28} (1978) 822;
%%CITATION = YAFIA,28,1597;%%
I.~I.~Balitsky and L.~N.~Lipatov, JETP\ Lett.\ \textbf{30} (1979) 355.
%%CITATION = ZFPRA,30,383;%%



\bibitem{KL00}
A.~V.~Kotikov and L.~N.~Lipatov, Nucl.\ Phys.\ \textbf{B582} (2000) 19.
%%CITATION = NUPHA,B582,19;%%

\bibitem{KL}
A.~V.~Kotikov and L.~N.~Lipatov, Nucl.\ Phys.\ \textbf{B661} (2003) 19;
%%CITATION = NUPHA,B661,19;%%
in: {\it Proc. of the XXXV
Winter School}, Repino, S'Peterburg, 2001 (hep-ph/0112346).
%arXiv:hep-ph/0112346.
%%CITATION = HEP-PH/0112346;%%

%\cite{Fadin:2007xy}
\bibitem{Fadin:2007xy}
  V.~S.~Fadin and R.~Fiore,
  %``The dipole form of the BFKL kernel in supersymmetric Yang--Mills
  %theories,''
  Phys.\ Lett.\  B {\bf 661} (2008) 139
  [arXiv:0712.3901 [hep-ph]];
  %%CITATION = PHLTA,B661,139;%%
%\cite{Fadin:2009gh}
%\bibitem{Fadin:2009gh}
  V.~S.~Fadin, R.~Fiore and A.~V.~Grabovsky,
  %``Matching of the low-x evolution kernels,''
  Nucl.\ Phys.\  B {\bf 831} (2010) 248
  [arXiv:0911.5617 [hep-ph]].\\
  %%CITATION = NUPHA,B831,248;%%
%\cite{Balitsky:2009yp}
%\bibitem{Balitsky:2009yp}
  I.~Balitsky and G.~A.~Chirilli,
  %``High-energy amplitudes in N=4 SYM in the next-to-leading order,''
  Phys.\ Lett.\ B {\bf 687} (2010) 204
  [arXiv:0911.5192 [hep-ph]].
  %%CITATION = ARXIV:0911.5192;%%

\bibitem{next}
V.~S.~Fadin and L.~N.~Lipatov, Phys.\ Lett.\ B \textbf{429} (1998) 127; \newline
G. Camici and M. Ciafaloni, Phys. Lett. \textbf{B430} (1998) 349.
%%CITATION = PHLTA,B429,127;%%
%G.~Camici and M.~Ciafaloni, Phys.\ Lett.\ B \textbf{430} (1998) 349.
%%%CITATION = PHLTA,B430,349;%%

\bibitem{L93} L.~N.~Lipatov, Phys.\ Lett.\ B \textbf{309} (1993) 394; 
\textit{High energy asymptotics of mult-color QCD and exactly solvable lattice models},
hep-th/9311037, unpublished.

\bibitem{L97} L.~N.~Lipatov, talk at "Perspectives in Hadronic Physics", 
\textit{Proc. of Conf. ICTP}, Triest, Italy, May 1997.

\bibitem{Lipatov00}  L.N. Lipatov, in: \textit{Proc. of the Int. Workshop on very
high multiplicity physics}, Dubna, 2000, pp.159-176;
L. N. Lipatov, Nucl. Phys. Proc. Suppl. \textbf{99A} (2001) 175;
%\cite{Dolan:2001tt}
%\bibitem{Dolan:2001tt}
  F.~A.~Dolan and H.~Osborn,
  %``Superconformal symmetry, correlation functions and the operator product
  %expansion,''
  Nucl.\ Phys.\  B {\bf 629} (2002) 3
  [arXiv:hep-th/0112251].
  %%CITATION = NUPHA,B629,3;%%

%\cite{Kotikov:2003fb}
\bibitem{Kotikov:2003fb}
  A.~V.~Kotikov, L.~N.~Lipatov and V.~N.~Velizhanin,
  %``Anomalous dimensions of Wilson operators in N = 4 SYM theory,''
  Phys.\ Lett.\  B {\bf 557} (2003) 114
  [arXiv:hep-ph/0301021].
  %%CITATION = PHLTA,B557,114;%%


%\cite{KLOV}
\bibitem{Kotikov:2004er}
  A.~V.~Kotikov, L.~N.~Lipatov, A.~I.~Onishchenko and V.~N.~Velizhanin,
  %``Three-loop universal anomalous dimension of the Wilson operators in N =  4
  %SUSY Yang-Mills model,''
  Phys.\ Lett.\  B {\bf 595} (2004) 521
%  [Erratum-ibid.\  B {\bf 632} (2006) 754]
  [arXiv:hep-th/0404092].
  %%CITATION = PHLTA,B595,521;%%


%\cite{Kotikov:2007cy}
\bibitem{Kotikov:2007cy}
  A.~V.~Kotikov, L.~N.~Lipatov, A.~Rej, M.~Staudacher and V.~N.~Velizhanin,
  %``Dressing and Wrapping,''
  J.\ Stat.\ Mech.\  {\bf 0710} (2007) P10003
  [arXiv:0704.3586 [hep-th]].
  %%CITATION = JSTAT,0710,P10003;%%
%\cite{Bajnok:2008qj}
%\bibitem{Bajnok:2008qj}

\bibitem{Janik}
 Z.~Bajnok, R.~A.~Janik and T.~Lukowski,
  %``Four loop twist two, BFKL, wrapping and strings,''
  Nucl.\ Phys.\  B {\bf 816} (2009) 376
  [arXiv:0811.4448 [hep-th]].
  %%CITATION = NUPHA,B816,376;%%


%\cite{Lukowski:2009ce}
\bibitem{Lukowski:2009ce}
  T.~Lukowski, A.~Rej and V.~N.~Velizhanin,
  %``Five-Loop Anomalous Dimension of Twist-Two Operators,''
  Nucl.\ Phys.\  B {\bf 831} (2010) 105
  [arXiv:0912.1624 [hep-th]].
  %%CITATION = NUPHA,B831,105;%%


%\cite{Moch:2004pa}
\bibitem{Moch:2004pa}
  S.~Moch, J.~A.~M.~Vermaseren and A.~Vogt,
  %``The three-loop splitting functions in QCD: The non-singlet case,''
  Nucl.\ Phys.\  B {\bf 688} (2004) 101
  [arXiv:hep-ph/0403192].
  %%CITATION = NUPHA,B688,101;%%

%\cite{Beisert:2005fw}
\bibitem{Beisert:2005fw}
  N.~Beisert and M.~Staudacher,
  %``Long-range PSU(2,2|4) Bethe ansaetze for gauge theory and strings,''
  Nucl.\ Phys.\  B {\bf 727} (2005) 1
  [arXiv:hep-th/0504190];
  %%CITATION = NUPHA,B727,1;%%
%\cite{Staudacher:2004tk}
%\bibitem{Staudacher:2004tk}
  M.~Staudacher,
  %``The factorized S-matrix of CFT/AdS,''
  JHEP {\bf 0505} (2005) 054
  [arXiv:hep-th/0412188].
  %%CITATION = JHEPA,0505,054;%%



%\cite{Kotikov:2008pv}
\bibitem{Kotikov:2008pv}
  A.~V.~Kotikov, A.~Rej and S.~Zieme,
  %``Analytic three-loop Solutions for N=4 SYM Twist Operators,''
  Nucl.\ Phys.\  B {\bf 813} (2009) 460
  [arXiv:0810.0691 [hep-th]];
  %%CITATION = NUPHA,B813,460;%%
%\cite{Beccaria:2009rw}
%\bibitem{Beccaria:2009rw}
  M.~Beccaria, A.~V.~Belitsky, A.~V.~Kotikov and S.~Zieme,
  %``Analytic solution of the multiloop Baxter equation,''
  Nucl.\ Phys.\  B {\bf 827} (2010) 565
  [arXiv:0908.0520 [hep-th]].
  %%CITATION = NUPHA,B827,565;%%


\bibitem{AdS-CFT}  J.~Maldacena, Adv.\ Theor.\ Math.\ Phys.\ \textbf{2}
(1998) 231; Int.\ J.\ Theor.\ Phys.\ \textbf{38} (1998) 1113;
%%CITATION = IJTPB,38,1113;%%

\bibitem{AdS-CFT1}
S.~S.~Gubser, I.~R.~Klebanov and A.~M.~Polyakov, Phys.\
Lett.\ B \textbf{428} (1998) 105.
%%CITATION = PHLTA,B428,105;%%
%E.~Witten, Adv.\ Theor.\ Math.\ Phys.\
%\textbf{2} (1998) 253.
%%%CITATION = 00203,2,253;%%

\bibitem{Witten}
E.~Witten, Adv.\ Theor.\ Math.\ Phys.\
\textbf{2} (1998) 253.
%%CITATION = 00203,2,253;%%


%\cite{Beisert:2006ez}
\bibitem{Beisert:2006ez}
  N.~Beisert, B.~Eden and M.~Staudacher,
  %``Transcendentality and crossing,''
  J.\ Stat.\ Mech.\  {\bf 0701} (2007) P021
  [arXiv:hep-th/0610251].
  %%CITATION = JSTAT,0701,P021;%%

%\cite{Benna:2006nd}
\bibitem{Benna:2006nd}
  M.~K.~Benna, S.~Benvenuti, I.~R.~Klebanov and A.~Scardicchio,
  %``A test of the AdS/CFT correspondence using high-spin operators,''
  Phys.\ Rev.\ Lett.\  {\bf 98} (2007) 131603
  [arXiv:hep-th/0611135];
  %%CITATION = PRLTA,98,131603;%%
%\cite{Kotikov:2006ts}
%\bibitem{Kotikov:2006ts}
  A.~V.~Kotikov and L.~N.~Lipatov,
  %``On the highest transcendentality in N = 4 SUSY,''
  Nucl.\ Phys.\  B {\bf 769} (2007) 217
  [arXiv:hep-th/0611204].
  %%CITATION = NUPHA,B769,217;%%

%\cite{Basso:2007wd}
\bibitem{Basso:2007wd}
  B.~Basso, G.~P.~Korchemsky and J.~Kotanski,
  %``Cusp anomalous dimension in maximally supersymmetric Yang-Mills theory at
  %strong coupling,''
  Phys.\ Rev.\ Lett.\  {\bf 100} (2008) 091601
  [arXiv:0708.3933 [hep-th]].
  %%CITATION = PRLTA,100,091601;%%


%\cite{Gubser:2002tv}
\bibitem{Gubser:2002tv}
  S.~S.~Gubser, I.~R.~Klebanov and A.~M.~Polyakov,
  %``A semi-classical limit of the gauge/string correspondence,''
  Nucl.\ Phys.\  B {\bf 636} (2002) 99
  [arXiv:hep-th/0204051].
  %%CITATION = NUPHA,B636,99;%%

%\cite{Frolov:2002av}
\bibitem{Frolov:2002av}
  S.~Frolov and A.~A.~Tseytlin,
  %``Semiclassical quantization of rotating superstring in AdS(5) x S(5),''
  JHEP {\bf 0206} (2002) 007
  [arXiv:hep-th/0204226];
  %%CITATION = JHEPA,0206,007;%%
%\cite{Roiban:2007jf}
%\bibitem{Roiban:2007jf}
  R.~Roiban, A.~Tirziu and A.~A.~Tseytlin,
  %``Two-loop world-sheet corrections in AdS_5 x S^5 superstring,''
  JHEP {\bf 0707} (2007) 056
  [arXiv:0704.3638 [hep-th]].
  %%CITATION = JHEPA,0707,056;%%



\bibitem{Polchinski:2002jw}
%\cite{Polchinski:2001tt}
%\bibitem{Polchinski:2001tt}
  J.~Polchinski and M.~J.~Strassler,
  %``Hard scattering and gauge / string duality,''
  Phys.\ Rev.\ Lett.\  {\bf 88} (2002) 031601
  [arXiv:hep-th/0109174];
  %%CITATION = PRLTA,88,031601;%%
%\bibitem{Polchinski:2002jw}
%  J.~Polchinski and M.~J.~Strassler,
  %``Deep inelastic scattering and gauge/string duality,''
  JHEP {\bf 0305} (2003) 012
  [arXiv:hep-th/0209211].
  %%CITATION = JHEPA,0305,012;%%

\bibitem{Kotikov:2006}
 A.~V.~Kotikov, L.~N.~Lipatov, A.~I.~Onishchenko and V.~N.~Velizhanin,
Phys.\ Lett.\  B {\bf 632} (2006) 754
[arXiv:hep-th/0404092v5].


%\cite{Brower:2006ea}
\bibitem{Brower:2006ea}
  R.~C.~Brower, J.~Polchinski, M.~J.~Strassler and C.~I.~Tan,
  %``The Pomeron and Gauge/String Duality,''
  JHEP {\bf 0712} (2007) 005
  [arXiv:hep-th/0603115].
  %%CITATION = JHEPA,0712,005;%%


%\cite{Stasto:2007uv}
\bibitem{Stasto:2007uv}
  A.~M.~Stasto,
  %``The BFKL Pomeron in the weak and strong coupling limits and kinematical
  %constraints,''
  Phys.\ Rev.\  D {\bf 75} (2007) 054023
  [arXiv:hep-ph/0702195].
  %%CITATION = PHRVA,D75,054023;%%

%\cite{Lipatov:2011ab}
\bibitem{Lipatov:2011ab}
  L.~N.~Lipatov,
  %``Effective action for the Regge processes in gravity,''
  arXiv:1105.3127 [hep-th], Part. Nucl. Phys. (2013) in press.
  %%CITATION = ARXIV:1105.3127;%%


%\cite{Gromov:2011de}
\bibitem{Gromov:2011de}
  N.~Gromov, D.~Serban, I.~Shenderovich and D.~Volin,
  %``Quantum folded string and integrability: from finite size effects to
  %Konishi dimension,''
  JHEP {\bf 1108} (2011) 046
  [arXiv:1102.1040 [hep-th]].
  %%CITATION = JHEPA,1108,046;%%


%\cite{Basso:2011rs}
\bibitem{Basso:2011rs}
  B.~Basso,
  %``An exact slope for AdS/CFT,''
  arXiv:1109.3154 [hep-th];
  %%CITATION = ARXIV:1109.3154;%%
%\cite{Basso:2012ex}
%\bibitem{Basso:2012ex}  B.~Basso,
  %``Scaling dimensions at small spin in N=4 SYM theory,''
  arXiv:1205.0054 [hep-th].
  %%CITATION = ARXIV:1205.0054;%%

%\cite{Gromov:2011bz}
\bibitem{Gromov:2011bz}
  N.~Gromov and S.~Valatka,
  %``Deeper Look into Short Strings,''
  JHEP {\bf 1203} (2012) 058
  [arXiv:1109.6305 [hep-th]].
  %%CITATION = ARXIV:1109.6305;%%



%\cite{Roiban:2011fe}
\bibitem{Roiban:2011fe}
  R.~Roiban and A.~A.~Tseytlin,
  %``Semiclassical string computation of strong-coupling corrections to
  %dimensions of operators in Konishi multiplet,''
  Nucl.\ Phys.\  B {\bf 848} (2011) 251
  [arXiv:1102.1209 [hep-th]].
  %%CITATION = NUPHA,B848,251;%%


\bibitem{Lewin}  L. Lewin, Polylogarithms and Associated Functions (North
Holland, Amsterdam, 1981).



\bibitem{DGLAP}
V.~N.~Gribov and L.~N.~Lipatov, Sov.\ J.\ Nucl.\ Phys.\ \textbf{15} (1972) 438;
%%CITATION = YAFIA,15,781;%%
V.~N.~Gribov and L.~N.~Lipatov, Sov.\ J.\ Nucl.\ Phys.\ \textbf{15} (1972) 675;
%%CITATION = YAFIA,15,1218;%%
L.~N.~Lipatov, Sov.\ J.\ Nucl.\ Phys.\ \textbf{20} (1975) 94;
%%CITATION = YAFIA,20,181;%%
G.~Altarelli and G.~Parisi, Nucl.\ Phys.\ \textbf{B126} (1977) 298;
%%CITATION = NUPHA,B126,298;%%
Yu.~L. Dokshitzer, Sov.\ Phys.\ JETP \textbf{46} (1977) 641.
%%CITATION = ZETFA,73,1216;%%

%\cite{Salam:1998tj}
\bibitem{Salam:1998tj}
  G.~P.~Salam,
  %``A resummation of large sub-leading corrections at small x,''
  JHEP {\bf 9807} (1998) 019
  [arXiv:hep-ph/9806482].
  %%CITATION = JHEPA,9807,019;%%
%\cite{Salam:1999cn}
%\bibitem{Salam:1999cn}
%  G.~P.~Salam,
  %``An introduction to leading and next-to-leading BFKL,''
  Acta Phys.\ Polon.\  B {\bf 30} (1999) 3679
  [arXiv:hep-ph/9910492].
  %%CITATION = APPOA,B30,3679;%%




%\cite{Brower:2007xg}
\bibitem{Brower:2007xg}
  R.~C.~Brower, M.~J.~Strassler and C.~I.~Tan,
  %``On The Pomeron at Large 't Hooft Coupling,''
  JHEP {\bf 0903} (2009) 092
  [arXiv:0710.4378 [hep-th]];
  %%CITATION = JHEPA,0903,092;%%
%\cite{Brower:2008cy}
%\bibitem{Brower:2008cy}
  R.~C.~Brower, M.~Djuric and C.~I.~Tan,
  %``Odderon In Gauge/String Duality,''
  JHEP {\bf 0907} (2009) 063
  [arXiv:0812.0354 [hep-th]].
  %%CITATION = JHEPA,0907,063;%%

%\cite{Costa:2012cb}
\bibitem{Costa:2012cb}
  M.~S.~Costa, V.~Goncalves and J.~Penedones,
  %``Conformal Regge theory,''
  JHEP {\bf 2012} (2012) 091
  [arXiv:1209.4355 [hep-th]].
  %%CITATION = ARXIV:1209.4355;%%

%\bibitem{VelizhComm}
%V.~N.~Velizhanin, private communications

%%\cite{Bern:2006ew}
\bibitem{Bern:2006ew}
  Z.~Bern, M.~Czakon, L.~J.~Dixon, D.~A.~Kosower and V.~A.~Smirnov,
%  %``The Four-Loop Planar Amplitude and Cusp Anomalous Dimension in Maximally
%  %Supersymmetric Yang-Mills Theory,''
Phys.\ Rev.\  D {\bf 75} (2007) 085010
  [arXiv:hep-th/0610248].
%  %%CITATION = PHRVA,D75,085010;%%

\bibitem{VelizhComm}
V.~N.~Velizhanin, private communications


%\cite{Beccaria:2012tu}
\bibitem{Beccaria:2012tu}
  M.~Beccaria and G.~Macorini,
  %``Resummation of semiclassical short folded string,''
  JHEP {\bf 1202} (2012) 092
  [arXiv:1201.0608 [hep-th]].
  %%CITATION = ARXIV:1201.0608;%%

%%\cite{Tseytlin:2008gb}
%\bibitem{Tseytlin:2008gb}
%  A.~A.~Tseytlin,
%  %``Introductory Lectures on String Theory,''
%  arXiv:0808.0663 [physics.pop-ph].
%  %%CITATION = ARXIV:0808.0663;%%




%\cite{Tirziu:2008fk}
\bibitem{Tirziu:2008fk}
  A.~Tirziu and A.~A.~Tseytlin,
  %``Quantum corrections to energy of short spinning string in AdS5,''
  Phys.\ Rev.\  D {\bf 78} (2008) 066002
  [arXiv:0806.4758 [hep-th]];
  %%CITATION = PHRVA,D78,066002;%%
%\cite{Beccaria:2010ry}
%\bibitem{Beccaria:2010ry}
  M.~Beccaria, G.~V.~Dunne, V.~Forini, M.~Pawellek and A.~A.~Tseytlin,
  %``Exact computation of one-loop correction to energy of spinning folded
  %string in AdS_5 x S^5,''
  J.\ Phys.\ A {\bf 43} (2010) 165402
  [arXiv:1001.4018 [hep-th]];
  %%CITATION = JPAGB,43,165402;%%
%\cite{Beccaria:2010zn}
%\bibitem{Beccaria:2010zn}
  M.~Beccaria, G.~V.~Dunne, G.~Macorini, A.~Tirziu and A.~A.~Tseytlin,
  %``Exact computation of one-loop correction to energy of pulsating strings in
  %AdS_5 x S^5,''
  J.\ Phys.\ A  {\bf 44} (2011) 015404
  [arXiv:1009.2318 [hep-th]];
  %%CITATION = JPAGB,A44,015404;%%
%\cite{Roiban:2009aa}
%\bibitem{Roiban:2009aa}
  R.~Roiban and A.~A.~Tseytlin,
  %``Quantum strings in AdS_5 x S^5: strong-coupling corrections to dimension of
  %Konishi operator,''
  JHEP {\bf 0911} (2009) 013
  [arXiv:0906.4294 [hep-th]].



%\cite{Giombi:2010zi}
%\bibitem{Giombi:2010zi}
%  S.~Giombi, R.~Ricci, R.~Roiban and A.~A.~Tseytlin,
%  %``Two-loop AdS_5 x S^5 superstring: testing asymptotic Bethe ansatz and
%  %finite size corrections,''
%  arXiv:1010.4594 [hep-th].
%  %%CITATION = ARXIV:1010.4594;%%


%\cite{Gromov:2009zb}
\bibitem{Gromov:2009zb}
  N.~Gromov, V.~Kazakov and P.~Vieira,
  %``Exact Spectrum of Planar ${\cal N}=4$ Supersymmetric Yang-Mills Theory:
  %Konishi Dimension at Any Coupling,''
  Phys.\ Rev.\ Lett.\  {\bf 104} (2010) 211601
  [arXiv:0906.4240 [hep-th]].
  %%CITATION = PRLTA,104,211601;%%


%\cite{Frolov:2012zv}
\bibitem{Frolov:2012zv}
  S.~Frolov,
  %``Scaling dimensions from the mirror TBA,''
  J.\ Phys.\ A {\bf 45}, 305402 (2012)
  [arXiv:1201.2317 [hep-th]].
  %%CITATION = ARXIV:1201.2317;%%


\end{thebibliography}
\end{document}